\documentclass[twocolumn,prl,showpacs,superscriptaddress,preprintnumbers]{revtex4}
\usepackage{amsmath,amssymb,epsfig,color}

\begin{document}

\title{Formation of an unconventional Ag valence state in Ag$_2$NiO$_2$.}
\author{M.D. Johannes}
\affiliation{Code 6393, Naval Research Laboratory, Washington, D.C. 20375}
\author{S. Streltsov}
\affiliation{Institute of Metal Physics, S.Kovalevskoy St. 18, 620219 Ekaterinburg GSP-170, Russia}
\affiliation{II. Physikalisches Institut, Universit$\ddot a$t zu K$\ddot o$ln,
Z$\ddot u$lpicher Stra$\ss$e 77, D-50937 K$\ddot o$ln, Germany}
\author{I.I. Mazin}
\affiliation{Code 6393, Naval Research Laboratory, Washington, D.C. 20375}
\author{D.I. Khomskii}
\affiliation{II. Physikalisches Institut, Universit$\ddot a$t zu K$\ddot o$ln,
Z$\ddot u$lpicher Stra$\ss$e 77, D-50937 K$\ddot o$ln, Germany}

\begin{abstract} The Ag ion in the recently synthesized novel material Ag$_2$NiO$_2$ adopts an extremely unusual valency of
$\frac{1}{2}$, leaving the Ni ion as 3$^+$, rather than the expected 2$^+$.  Using first principles calculations, we show
that this mysterious subvalent state emerges due to a strong bonding-antibonding interaction between the two Ag layers
which drives the lower band beneath the O $p$ complex, eliminating the possibility of a conventional Ag $1^+$ valence
state.  The strong renormalization of the specific heat coefficient, $\gamma$, is likely due to strong spin fluctuations
that stem from nearly complete compensation of the ferro- (metallic double exchange and the 90$^\circ$ superexchange) and
antiferromagnetic (conventional superexchange via Ni-O-Ag-O-Ni path) interactions. \end{abstract}
   
\maketitle

As with other noble metals, Ag typically adopts a formal valence of 1, 2 or 3 in a compound, corresponding
 to an empty $s$-shell and either a full or partially depleted $d$-shell.  Any
valence less than 1 would leave the Ag $s$-shell partially filled and is therefore rather unnatural, particularly with
strong oxidizers such as O or F.  To our knowledge, there is but one case in which Ag assumes a formal
valency of $\frac{1}{2}$:  Ag$_{2}$F \cite{Williams-88,exc}, and this compound is rather unstable: it
 decomposes in the presence of water,
ultraviolet light, or above 80$^{\circ }$C.

Since F$^{2-}$ cannot exist in nature, Ag in Ag$_{2}$F is perforce $\frac{1}{2}^+$.  Ni$^{2+}$, on the
other hand, is a common oxidation state for nickel, just as Ag$^{+}$ is for silver and one would therefore
anticipate that the Ag$ _{2}$NiO$_{2}$ compound should form, with Ag$^{+}$ and Ni$^{2+}$ oxidation
states, and thus be a magnetic insulator, like NiO.
Recently  Ag$_{2}$NiO$_{2},$ has been
synthesized~\cite{Schreyer-02}, and, intriguingly, did not fit this picture~\cite{Schreyer-02,Yoshida-06}.  
It remained metallic down to the lowest measured temperatures,
with properties rather close to those of delaffosite AgNiO$_{2}$, a nearly-isostructural compound with one less silver 
\cite{dela}. This
has been interpreted as a signal that Ag is underoxidized, or subvalent [Ag$_{2}$]$^{+}$, leaving Ni in a $3+$ state
\cite{Schreyer-02}. In such a case it would indeed be similar to AgNiO$_{2}$ as well as to better known nickelates,
LiNiO$_{2}$ or NaNiO$_{2}$, with the low-spin ($t_{2g}^{6}e_{g}^{1}$) Ni$^{3+}$ being a Jahn-Teller (JT) ion. Not
surprisingly then, just as in NaNiO$_{2},$ a structural transition from a high temperature rhombohedral phase to a
(presumably) monoclinic phase was reported in Ag$_{2}$NiO$_{2}$ at $T_{s}$=260~K~\cite{Schreyer-02,Yoshida-06,Jansen-06}.
This was attributed in Ref. \onlinecite{Yoshida-06} to a cooperative Jahn-Teller transition of Ni$^{3+}$, accompanied by
orbital ordering (OO), although this attribution was later questioned\cite{Sugiyama-06}. 
%However, other compounds based on intercalated [NiO$_{2}]^{-}$ layers show only a rather 
%weak tendency to or even absence of JT distortion (indicating that $d-$electrons there 
%are only moderately localized) and in 3D oxides Ni$^{3+}$ often shows no JT distortion 
%at all ($e.g.,$LaNiO$_{3}$). {\bf Is the point of this last sentence to emphasize that a 
%JT distorted state is surprising?}

This compound poses one principal puzzle: why does Ag assume such an unnatural valence state instead of the expected
combination of Ag$^{1+}$ and Ni$^{2+}$? Additionally, there are unexplained phenomena such as the large electronic
specific heat coefficient, $\gamma =19$ mJ/mole K$^{2}$ \cite{Yoshida-06}, and the undetermined nature of the magnetic
ordering. The layered nickelates and cobaltates, if magnetic, usually show an A-type antiferromagnetism (AFM) setting in at
low temperature. Ag$_{2}$NiO$_{2}$ has a N\'eel temperature of 56~K\cite{Schreyer-02}, but the ordering type is unknown,
and there are speculations\cite{Sugiyama-06} that it is not A-type, but rather AFM in-plane and possibly incommensurate.  
Unlike most similar materials, the Curie-Weiss temperature changes sign at the structural transition
$T_{s}=260$~K~\cite{Schreyer-02}, with the high-temperature undistorted phase showing net FM spin fluctuations ($\Theta
=10$~K), and the low-temperature phase exhibiting AFM ones ($\Theta =-30$~K), both lower than the observed ordering
temperature ($\vert \Theta |<T_{N}$).

In this Letter we shall demonstrate, using first principles calculations, that most, if not all of these puzzles can be
resolved on the one-electron level. The half-valent silver in this compound appears due to the strong
bonding-antibonding splitting in the Ag bilayer, pushing most of the bonding Ag-$s$ band below the Ni-$d$ complex and below
most of the O-$p$ states. The magnetic disparity with the other layered nickelates and cobaltates can be explained by the
fact that hopping between orthogonal O-$p$ orbitals on the same site becomes possible via the metallic Ag layers, thus
creating a path for an AFM exchange that compensates the usual metallic FM double exchange and the 90$^\circ$
FM superexchange via oxygen. This compensation leads to strong spin fluctuations
which may be responsible for the large specific heat.

The high-temperature crystal structure of Ag$_{2}$NiO$_{2}$ is rhombohedral 
($R\overline{3}m$), consisting of close packed triangular
layers of O-Ni-O-Ag-Ag-O-Ni-O, with edge-sharing NiO$_{6}$ octahedra. The stacking sequence is ABCABC so that the Ni layers
form a BACBAC sequence. For this phase, we used the structural parameters from Ref.~\cite{Schreyer-02}. We also performed
calculations in the monoclinic ($C2/m$) structure, similar to NaNiO$_{2}$, suggested by Jansen~\cite{Jansen-06}. We used a
full-potential augmented plane-wave with local orbitals (APW+lo) method~\cite{Wien2k}. To ensure complete convergence we used
up to 1024 inequivalent \textbf{k}-points with $RK_{\max }=7$. For magnetic supercell calculations, described in detail later,
we used up to 140 \textbf{k}-points in the (now four times smaller) Brillouin zone and $RK_{\max }$ up to 9. Gradient
corrections to the exchange correlation potential were included in form of Ref.\cite{PBE}. The calculated band structure
appears to correspond to the Ag$^{0.5+}-$Ni$^{3+}$ combination, as conjectured by Jensen \textit{et. 
al.} \cite{Schreyer-02}.
To understand the microscopic reasons for such an unusual result, we employed an orbital downfolding procedure as
built into the linear muffin-tin orbital (LMTO) method in the atomic spheres approximation (ASA) \cite{Andersen-84}.

\begin{figure}[tbp]
\includegraphics[clip=false,width=0.47\textwidth]{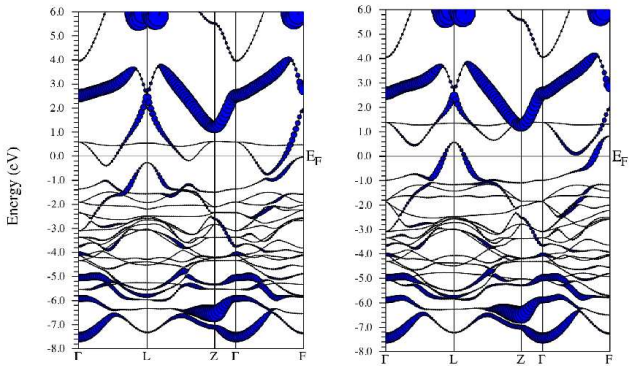}
\caption{(color online) LAPW band structure obtained with GGA for minority
(left panel) and majority (right panel) spins. Ag$-s$ character is
emphasized by fat blue circles.}
\label{LAPW-bands}
\end{figure}

We begin by investigating the high-temperature structure with FM
ordering. The LAPW bands with the Ag-$s$ character emphasized are shown
in Fig.~\ref{LAPW-bands}. One can see a nearly free electron band starting
at $\sim -8$ eV at the $\Gamma $ point, then lost in the manifold
of the O$-p$ and Ag$-d$ bands between $\sim -7$ and $\sim -2$eV, 
and reappearing above the Fermi level at L and F. Surprisingly,
the bottom of this band (at $\Gamma$) lies $below$ the bottom
of the  O$-p$ bands (at L). This band originates from the Ag-$s$  and
$p$ states, and, since it starts below the O$-p$ states, the
valence of Ag, no matter how one chooses to define it, must be less than
1. The obvious question is how can the conductivity band of an $s$-metal
drop below O$-p$ states? To answer this question, we separate the band
structure into individual contributions, using the powerful band
downfolding technique of the LMTO method.

\begin{figure}[tbp]
\centering
\includegraphics[clip=false,width=0.47\textwidth]{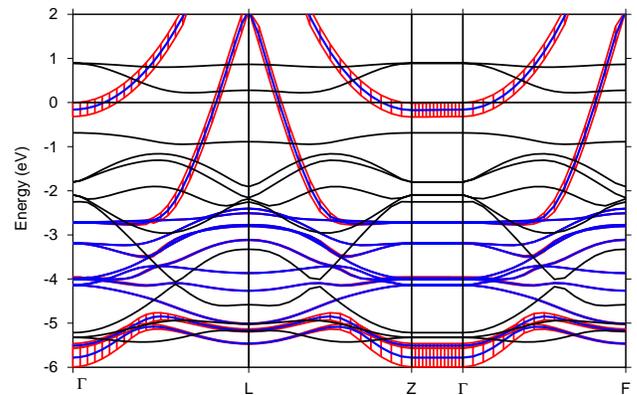}
\caption{(color online) Downfolded LMTO band structure obtained within LDA:
NiO (black,solid), Ag-$s,p,d$ (blue, thick) bands. The contribution of Ag-$%
s,p$ bands is shown as red fatbands.}
\label{LMTO-bands}
\end{figure}

Fig.\ref{LMTO-bands}a shows the LMTO bands,
%\cite{LMTO-details} 
obtained for the same structure, but, for simplicity, without
spin polarization. We observe that the qualitative character of the LAPW bands, including the relative positions of the Ag
and O bands, is reproduced. To understand this band structure, we first remove from the LMTO basis set all states except
Ni$-3d$ and O$-2p$. Note that the self-consistent potential for the real compound is still used, so the resulting band
structure is $not$ the same as for a hypothetical NiO$_{2}$ with Ag atom removed. It represents the actual Ni and O states
in Ag$_{2}$NiO$_{2}$, but with the hybridization with other bands switched off.  The bottom of the O$-2p$ bands is located
away from the $\Gamma -$point at $\sim $ $-5$ eV, and their top overlaps the Ni$-3d$ band around $-2$ eV. We
can also remove Ni$-3d$ and O$-2p$ from the basis, leaving only Ag $s,p$ and $d$. In this case, we clearly see two bonding 
and 
antibonding
$sp-$bands that are free-electron-like along in-plane directions (emphasized in Fig.~\ref{LMTO-bands} in red). The bonding
band intersects the narrow Ag$-d$ bands and finally reappears below the lowest O$-p$ band. The bonding-antibonding
splitting is very large, encompassing half the full band dispersion from $\Gamma -$point to the edge of the Brillouin zone.
This can be understood from the fact that each Ag has 6 nearest neighbors in its own plane compared to 3 in
the neighboring plane. 
%Another way to visualize this is to imagine $fcc$ Ag without its two neighboring (111) planes.  Such
%a structure, after a small distortion, forms the Ag bilayer in our compound. Similarly NiO$_{2}$ trilayers can be visualized
%by removing three successive (111) planes in cubic NiO.

The bonding-antibonding splitting is the principal feature that distinguishes Ag$_{2}$NiO$_{2}$ from the similar one-Ag-layer
delafossite AgNiO$_{2}$. The latter behaves more like other $A$NiO$_{2}$-based layered materials (A=alkaline) intercalated
with a monovalent metal, such as LiNiO$_{2}$ or NaNiO$_{2}$.  However, even in the delafossite compound, the Ag$-s$ band is
considerably lower than the alkaline $s-$bands in $A$NiO$_{2},$ such that the very bottom of the s-band drops well below the
Fermi level (assuring metalicity), but not below the bottom of the O bands. Obviously, it is the bilayer band splitting in
Ag$_{2}$NiO$_{2}$ that pushes the bonding Ag band down compared to AgNiO$_{2}$, thus preventing charge transfer from this band
to O. Note that in both silver nickelates the bonding Ag band crosses the Fermi level and spans the same energy range as the
bands of the NiO$_{2}$ complex, thus preventing formation of an insulator even after JT splitting. 
%In addition, the presence
%of Ag$-sp$ bands in the region around E$_F$ leads to an unavoidable metallic screening of the on-site Coulomb electronic
%correlations (in a similar system, a NiO monolayer on Ag, screening reduces the Hubbard $U$ by 2~eV~\cite{Hao}.

\begin{figure}
\includegraphics[width = 0.95 \linewidth]{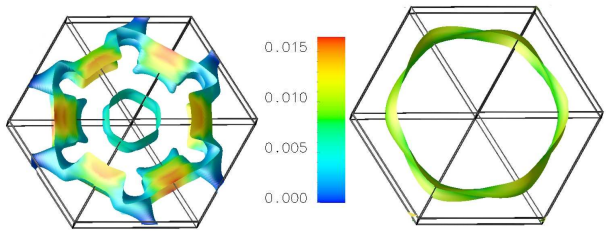}
\caption{(Color online) The Fermi surface of Ag$_2$NiO$_2$ in the high temperature (rhombohedral) phase in the 
spin-majority (a) and spin-minority (b) channel, colored according to their Fermi velocity (units are arbitrary).}
\label{FS}
\end{figure}

The combination of heavy $d$ and light $sp$ electrons in Ag$_{2}$NiO$_{2}$ should provide interesting transport properties. The
Fermi surface (FS) in this compound is comprised of a predominantly Ag$-sp,$ cylinder (Fig.\ref{FS}b), with faster electrons in the
spin minority channel (in this spin channel Ag$_{2}$NiO$_{2}$ is a 2D free electron metal) and a central hexagonal cylinder with a
surrounding network in the spin majority channel, formed by hybridized Ni$-d$ and Ag$-sp$ states. The characteristics of the FS in
the rhombohedral and monoclinic (to be discussed later) structures are displayed in Table\ref{TableDOS}.  Note that the density of
states is provided mostly by the Ni bands while conductivity is carried by the Ag bands. The former, though large, is a factor of 3
short of the experimental number (as estimated from the linear specific heat) in the rhombohedral cell and lessens by 
another a factor 
of two in the monoclinic cell. One cannot expect an electron-phonon renormalization
of this strength and therefore it must be of electronic origin.

\begin{table}
\caption{The plasma frequencies and density of states at E$_F$ for the high-temperature rhombohedral (R) and low 
temperature monoclinic (M) structures of 
Ag$_2$NiO$_2$ in the FM configuration.}
\begin{tabular}{|c|c|c|c|} \hline
& $\omega p_x (eV)$ &  $\omega p_z (eV) $ & $N(E_F) (eV^{-1})$ \\ \hline
R (up) & 2.83 & 2.58  & 2.24\\ \hline
R (dn) &2.96 & 1.12 & 0.25 \\ \hline
M (up) & 3.05 & 2.67 & 1.32\\ \hline
M (dn) &3.12 & 3.06 & 0.27\\ \hline
\end{tabular}
\label{TableDOS}
\end{table}

Generally, there are two effects of electronic correlation on the band structure. The first is due to the formation of
lower and upper Hubbard bands. It can be incorporated in a mean-field way in a method such as LDA+U \cite{Anisimov-91}.
However, this can only reduce the DOS at $E_{F}$ (N(E$_F$)), not enhance it, and in many correlated metallic systems, such as
CrO$_{2},$ Sr$ _{2}$RuO$_{4}$, or Na$_{x}$CoO$_{2}$, LDA+U actually worsens the agreement with experiment in terms of
magnetic properties or optical transition energies \cite{ldau}. We carried out LDA+U calculations (in the somewhat simpler
LMTO-ASA scheme) and indeed confirmed this conclusion.  Curiously, in LDA+U Ag$_2$NiO$_2$ remains metallic even with such 
a U as large as 8 eV.  The second effect is a narrowing of the quasiparticle peak near
E$_F$ that can be understood as dressing by low-lying bosonic excitations, such as soft spin fluctuations. This
effect enhances $\gamma $ and, generally speaking, depresses the magnetic moment. We shall argue that soft spin fluctuations
are indeed operative in Ag$_{2}$NiO$_{2}.$

The typical magnetic configuration for triangular oxide layers is A-type antiferromagnetism, or ferromagnetically aligned
planes stacked antiferromagnetically along the (111) rhombohedral direction.  The first surprise is that, contrary to the
alkali nickelates, the calculated interplane coupling is ferromagnetic (albeit by only 6 meV), but becomes effectively
degenerate (within 0.2 eV) upon the monoclinic JT distortion (a trend we found previously \cite{NaNiO2} in NaNiO$ _{2}$
where, however, the distortion is much stronger). This can be attributed to competition between the AFM interplane
superexchange and metallic FM double exchange. The distortion improves the superexchange paths \cite{NaNiO2} and 
slightly suppresses the FM double exchange through loss of DOS at $E_F$. 

\begin{figure} \includegraphics[width=0.95 \linewidth]{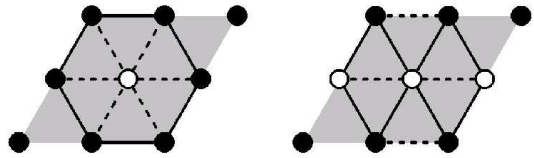} 
\caption{(Color online) Two of the three (FM not shown)
in-plane magnetic configurations investigated in this work.  One supercell is shown. Open (filled) circles show
 spin up (down) sites.  The twelve inrreducible bonds (FM and AFM) in each supercell are shown.} \label{mag}
\end{figure}

Next we consider the intraplane exchange. It has been argued\cite {Sugiyama-06} that the in-plane magnetism is itinerant (this
could explain why the Curie-Weiss temperature is small compared to $T_{N}$) and not FM (possibly incommensurate). To test this
computationally, we considered a 2x2x1 supercell (Fig.
\ref{mag}). Using a lowered symmetry to
allow for three inequivalent Ni sites, we created three magnetic patterns: $\mathcal{FM,FI}$ and $\mathcal{AF}$, each 
supercell containing 12 bonds and four spins.
In the $\mathcal{FI}$ pattern half the bonds are AFM and half FM, and three out of
four spins are parallel. In the $\mathcal{AF}$ pattern there are two up and two down spins, with 8 AFM and 4 FM bonds. 
In the
nearest neighbor approximation, $E_{\mathcal{AF}}-E_{\mathcal{FI}}=2J,$ where $J$ is the cost of changing one bond from FM to
AFM, and $E_{\mathcal{FI} }-E_{\mathcal{FM}}=6J.$ Our calculations give $E_{\mathcal{AF}}-E_{\mathcal{ FI}}=4.1$ meV, and
$E_{\mathcal{FI}}-E_{\mathcal{FM}}=15.0$ meV, consistent with a FM $J=2.3\pm 0.2$ meV. Note however that this number is very
small compared to typical triangular-layer nickelates and cobaltates. Moreover, in the rhombohedral structure the last energy
difference is reduced from 15 meV to less than 3, a full compensation (within computational accuracy)
of FM and AFM in-plane interaction. This can
again be traced to the effect of silver. Usually in-plane ferromagnetism is due to 90$^\circ$ superexchange: if
two transition metal ions and the bridging oxygen form a right triangle, then the corresponding oxygen orbitals are
orthogonal, oxygen-assisted hopping is suppressed, and the Hund's rule interaction at the O site provides an overall FM
exchange (see $e.g.$ Ref. \onlinecite{lno}). The Ni-O-Ni bond angle in Ag$_{2}$NiO$_{2}$ is indeed close to 90$ ^{\circ},$ but
the metallic Ag bands provide a channel for assisted hopping from one O orbital to another, creating an AFM superexchange
path: Ni-O-Ag-O-Ni (a similar mechanism was suggested for CuGeO$_2$ \cite{geert}). Competition between the two types of
superexchange results in a near compensation for the in-plane exchange, with the net sign depending on details beyond this
model\cite{note2}. 
%In addition to this nearest-neighbor coupling, longer range (RKKY-like) interactions 
%mediated by the metallic Ag$_2$ 
%layer may appear.

It is clear from the near cancellation of the FM and AFM interaction both in- and out-of-plane that Ag$_{2}$NiO$_{2}$ is a
perfect breeding ground for strong spin fluctuations. The ultimate ordering pattern, then, will depend on fine details of
the low-temperature structure. One of the main unanswered questions is whether this structure is JT distorted\cite
{Jansen-06} or not\cite{Sugiyama-06}. To address this question computationally we have performed a full structural
optimization using a pseudopotential code\cite{pp,note}. This approach has allowed us in the past to establish a tendency
to JT distortion in NaNiO$_{2}$ and LiNiO$_{2}$\cite{NaNiO2}. However, optimization in LDA, GGA or LDA+U invariably gave
the high-symmetry rhombohedral structure as the ground state. We have verified this result by calculating total energies
for the reported rhombohedral and monoclinic structures\cite{Jansen-06} in LAPW, and again found the former to be lower in
energy by 13 meV. While the energy differences involved are too small to definitively eliminate a JT-distorted ground
state, our first principles calculations lend support to the Sugiyama {\it et al} picture\cite{Sugiyama-06}.  In light of
the already substantial disagreement between calculated and measured $N(E_F)$, the loss of nearly half the DOS 
upon monoclinic distortion also casts doubt on the JT nature of the low temperature phase.

To summarize, we report all-electron, full-potential first principles calculations of the electronic, transport, magnetic
and structural properties of the novel exotic material Ag$_{2}$NiO$_{2}$. We confirm earlier conjectures\cite{Schreyer-02}
that the valence state of nickel in this compound is practically Ni$^{3+},$ while the valence of Ag is very close to
$\frac{1}{2}$. The microscopic reason for this unusual configuration appears to be the strong bonding-antibonding splitting
in the Ag$_{2}$ bilayer, which pushes the bottom of the Ag$-sp$ band below the O $-p$ bands. This results in the rather
unexpected $^{3+}$ Ni valence (because of high covalency in spectroscopic measurements this should correspond not to
$d^{7}$, but rather to $d^{8}L$ configuration of the ground state, where $L$ is a ligand hole). The Ag bilayer also affects
the magnetic properties non-trivially. On one hand, its metallic band provides an additional FM double exchange interaction
between the planes, offsetting the standard AFM superexchange, while on the other hand, overlap between the O-$p$ orbitals
and the metallic silver layer effectively removes the orthogonality of the O orbitals and revives an AFM superexchange,
compensating the usual in-plane ferromagnetism.  This creates strongly fluctuating spins that are responsible, in turn, for
a mass renormalization $m^{\ast }/m_{band}\approx 3.$ Finally, our first principles calculations cannot reproduce the
proposed\cite{Jansen-06} JT distortion, lending some support to the authors\cite{Sugiyama-06} who question this suggestion.
Overall, the physics of Ag$_{2}$NiO$_{2}$ appears to be unexpectedly rich and unusual and worth more detailed experimental
and theoretical study.

\textit{Acknowledgments} We would like to thank M.~Jansen for providing us
with their crystal structure data before publication; V. Anisimov and I.
Solovyev for fruitful duscussions. This work was supported by ONR and by the
projects INTAS 05-109-4727, RFFI-04-02-16096, and NWO 047.016.005, Ural
branch of RAS through the YS support program, and SFB 608. IIM wishes to
thank the university of Cologne for their hospitality during his sabbatical
visit.


\begin{thebibliography}{99}
\bibitem{exc} There exist a
few ternary and quaternary compounds where silver shows a formal valence
 slightly below 1, $e.g.,$ 4/5~\cite{Schreyer-02}, but
their valence assignment is more complicated.
\bibitem{note2} This effect is even clearer in AgNiO$_2$, where in-plane AFM order is
quite strong. Indeed, in AgNiO$_2$ silver sits right on top of an oxygen, providing an
easy hopping between two O-$p$ orbitals $via$ the same Ag.
\end{thebibliography}
\end{document}